\documentclass{aip-cp}

\usepackage[numbers]{natbib}
\usepackage{rotating}
\usepackage{graphicx}

\begin{document}

\title{The Laura++ Dalitz Plot Fitter}

\author[aff1]{Thomas Latham\corref{cor1}}

\affil[aff1]{Department of Physics, University of Warwick, Coventry, CV4 7AL, United Kingdom.}
\corresp[cor1]{Corresponding author: T.Latham@warwick.ac.uk}

\maketitle

\begin{abstract}
The Laura++ software package is designed for performing fits of amplitude models to data from decays of spin-0 particles into final states containing three spin-0 particles --- so-called Dalitz-plot analysis.
An overview of the amplitude formalism and its software implementation is given as well as summaries of recently added features and planned future developments.
\end{abstract}

\section{INTRODUCTION}

The Laura++ software package was originally developed by members of the University of Warwick particle physics group working on the BaBar experiment.
The group was working on amplitude analyses of the decays of $B$ mesons to three-body final states containing charmless pseudoscalar mesons, such as pions and kaons, i.e. Dalitz-plot analyses~\cite{Dalitz:1953cp}.
The Laura++ software was originally designed as a dedicated package for performing Dalitz-plot analyses, with the aim of being more straightforward to use but at the same time faster than the few packages that were then being used within the BaBar experiment.
The package was subsequently used in numerous publications from the BaBar collaboration, both amplitude analyses~\cite{Aubert:2005ce,Aubert:2005sk,Aubert:2008bj,Aubert:2009av,Aubert:2009me,delAmoSanchez:2010ad,Lees:2015uun} and other analyses of 3-body decays~\cite{Aubert:2007xb,Aubert:2008rr,Aubert:2008aw,delAmoSanchez:2010ur,BABAR:2011aaa}, from the LHCb collaboration~\cite{Aaij:2014xza,Aaij:2014baa,Aaij:2014afa,Aaij:2015vea,Aaij:2015kqa} and in some phenomenological works~\cite{Latham:2008zs,Gershon:2014yma}.
In September 2013 the code was publicly released on the HepForge website~\cite{HepForge} and now has numerous users within the BaBar, Belle, Belle II, LHCb, BES III and CMS collaborations.
The software consists of a set of C++ classes that are built into a library and a selection of example applications that link against that library.
The only external dependency is the ROOT scientific software framework~\cite{ROOT}.

\section{AMPLITUDE MODELS}

The majority of use cases of the package involve the formation of amplitude models for the decay of a spin-0 parent particle into three spin-0 children.
The two main modes of operation are then to use those models to either generate events using Monte Carlo methods or to fit the models to existing data (from experiment, simulation, etc.) in order to determine the parameters of interest.
One of the more commonly used amplitude models is the so-called isobar model~\cite{Fleming:1964zz,Morgan:1968zza,Herndon:1973yn}, which consists of describing the total decay amplitude by a sum of the contributing amplitudes, each with a complex coefficient:
\begin{equation}
	\mathcal{A}(x,y) = \sum_{j=1}^N c_j \, F_j(x,y)
\end{equation}
where $x$ and $y$ are the Dalitz-plot coordinates (i.e. they represent the position within the three-body phase space), $c_j$ is the complex coefficient for the contributing amplitude $j$ and $F_j(x,y)$ the corresponding dynamical function.
The $F_j$ functions are set within the code by choosing:
\begin{itemize}
	\item a particular model for the resonance (or nonresonant) lineshape, e.g. a relativistic Breit--Wigner,
	\item the two final-state particles to which the resonance decays,
	\item if appropriate, the mass, width and spin of the contributing resonance.
\end{itemize}
There are numerous models to choose from, as well as many pre-set values for the mass, width and spin of known resonances.
For example, one could use the following code snippet to define an isobar model for the decay $B^+ \to K^+ \pi^+ \pi^-$ containing three resonances, $K^*(892)^0$, $\rho(770)^0$ and $f_0(980)$, and a uniform nonresonant component:
\begin{verbatim}
// Define the parent and daughter particles
LauDaughters* daughters = new LauDaughters("B+", "K+", "pi+", "pi-");
// Define the veto regions in the Dalitz plot
LauVetoes* vetoes = new LauVetoes();
// Define the efficiency variation across the Dalitz plot
LauEffModel* effModel = new LauEffModel(daughters, vetoes);
// Define the dynamics and resonances
LauIsobarDynamics* sigModel = new LauIsobarDynamics(daughters, effModel);
LauAbsResonance* reson(0);
reson = sigModel->addResonance("K*0(892)",   2, LauAbsResonance::RelBW);
reson = sigModel->addResonance("rho0(770)",  1, LauAbsResonance::GS);
reson = sigModel->addResonance("f_0(980)",   1, LauAbsResonance::Flatte);
reson->setResonanceParameter("g1",0.2);
reson->setResonanceParameter("g2",1.0);
reson = sigModel->addResonance("NonReson",   0, LauAbsResonance::FlatNR);
\end{verbatim}
The first argument to the \texttt{addResonance} function identifies the resonance in order to obtain the pre-set values for the mass, width and spin.
The second argument indicates which of the final state particles (numbered from 1 to 3) is produced, together with the resonance, in the decay of the parent particle.
The third argument indicates the model to be used for this contribution.
Each of the contributions in this example uses a different amplitude model: relativistic Breit--Wigner, Gounaris--Sakurai~\cite{Gounaris:1968mw}, Flatt\'e~\cite{Flatte:1976xu}, uniform nonresonant.
One can also see an example of how to modify the values of model parameters, in this case the couplings to the $\pi\pi$ and $KK$ channels in the Flatt\'e model.

The complex coefficients are the main parameters of interest in most fits to data and must be expressed in terms of real numbers.
In the simplest case, they can be expressed either in terms of the real and imaginary part
\begin{equation}
\label{eq:realimag}
	c_j \equiv x_j + i \, y_j \,,
\end{equation}
or of the magnitude and phase
\begin{equation}
\label{eq:magphase}
	c_j \equiv a_j \, e^{i \theta_j} \,.
\end{equation}
However, the code to provide the complex coefficients to the model is essentially the same in all cases, except for the ``concrete'' type that is instantiated.
For example, using the form from Eq.~\ref{eq:magphase} for the resonances defined in the example above:
\begin{verbatim}
// Define the fitting model
LauSimpleFitModel* fitModel = new LauSimpleFitModel(sigModel);
// Create the coefficients for each contribution to the model
std::vector<LauAbsCoeffSet*> coeffset;
coeffset.push_back( new LauMagPhaseCoeffSet("K*0(892)",   1.00,  0.00,  kTRUE,  kTRUE) );
coeffset.push_back( new LauMagPhaseCoeffSet("rho0(770)",  0.53,  1.39, kFALSE, kFALSE) );
coeffset.push_back( new LauMagPhaseCoeffSet("f_0(980)",   0.27, -1.59, kFALSE, kFALSE) );
coeffset.push_back( new LauMagPhaseCoeffSet("NonReson",   0.54, -0.84, kFALSE, kFALSE) );
// Supply the coefficients to the model
std::vector<LauAbsCoeffSet*>::iterator iter=coeffset.begin();
for ( ; iter!=coeffset.end(); ++iter) {
    fitModel->setAmpCoeffSet(*iter);
}
\end{verbatim}
Note that one component (the $K^*(892)^0$ in the above example) must act as the reference amplitude, i.e. it must have its coefficient value fixed.
Furthermore, it is often chosen that this reference amplitude should be real and of unit magnitude, as is the case in the above example.
When considering decays of particle and antiparticle and the possible effects of $C\!P$ violation that might arise, the parameterisations can become more complicated.
Several such parameterisations are provided by the package.

\begin{figure}[tb]
  \centerline{
  \includegraphics[width=0.45\textwidth]{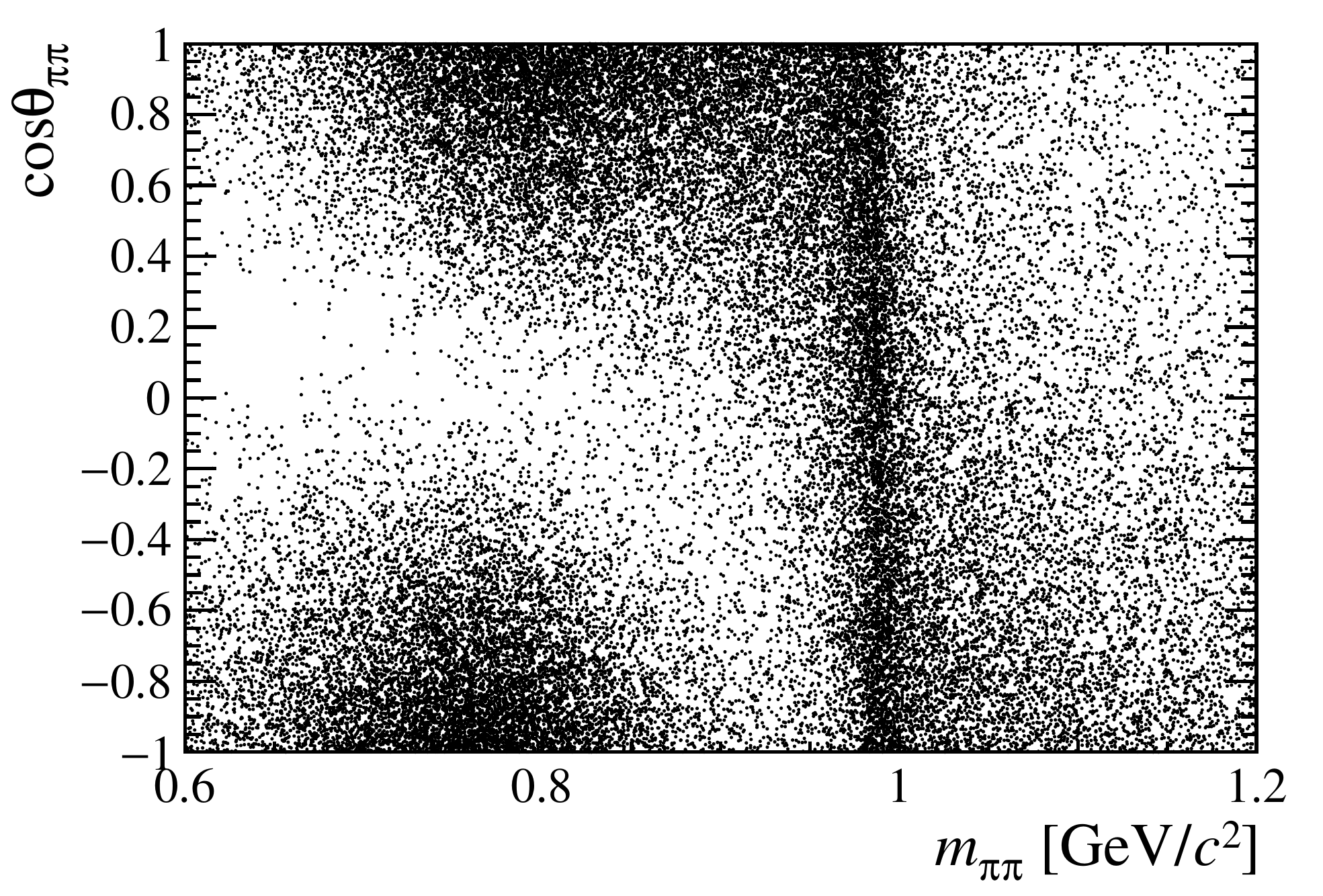}
  \includegraphics[width=0.45\textwidth]{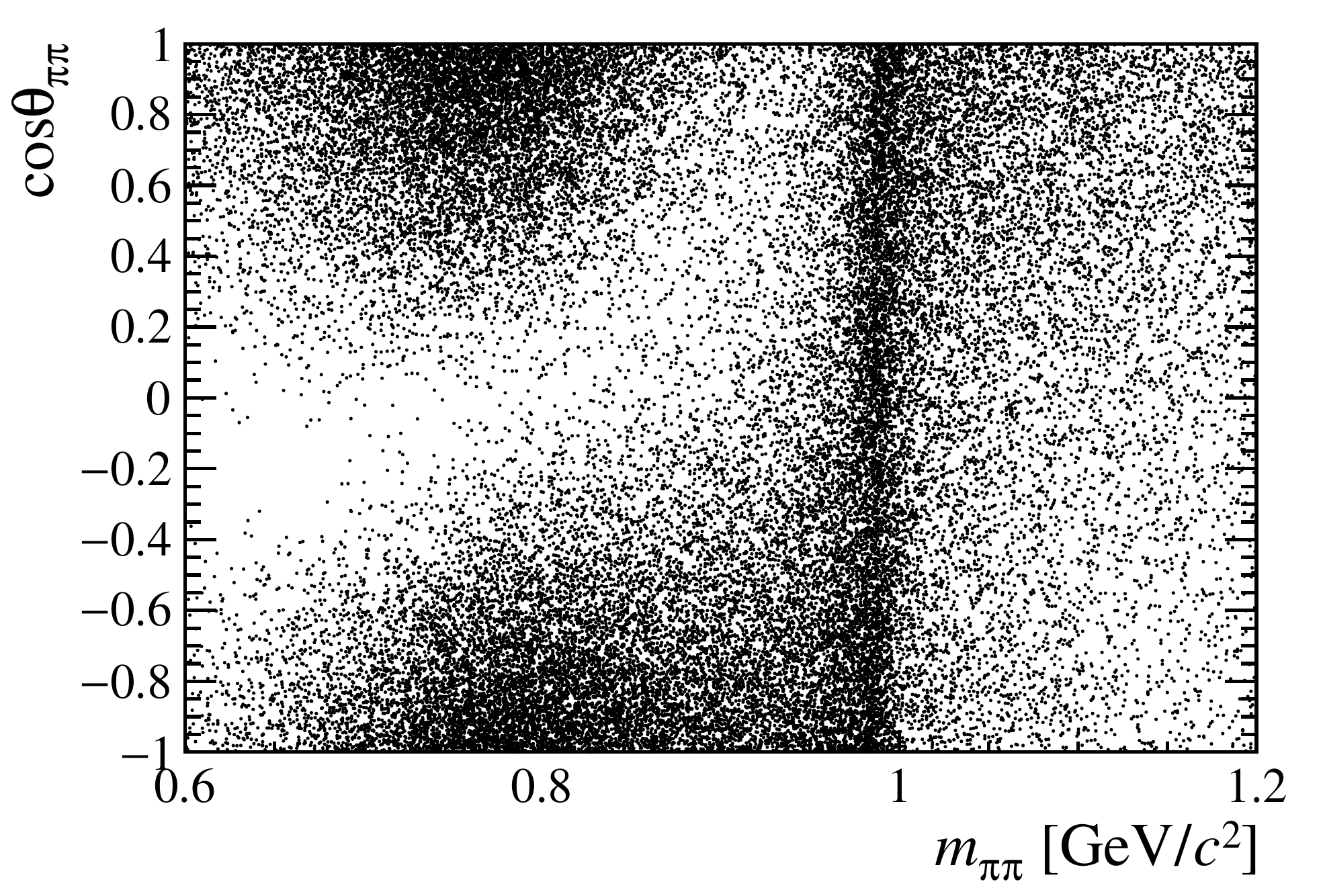}
  }
  \caption{Events generated by Laura++ in the $B^+ \to K^+ \pi^+ \pi^-$ Dalitz plot, where the model contains only the $\rho(770)^0$ and $f_0(980)$ resonances.  The magnitudes of the two resonances are the same, while their relative phase is either (left) $0^\circ$ or (right) $180^\circ$.}
  \label{fig:SP-interference}
\end{figure}

Figure~\ref{fig:SP-interference} shows the results of running Laura++ in generator mode for a simplified model containing just the $\rho(770)^0$ and $f_0(980)$ resonances for different values of the relative phase of the two resonances.
The events are plotted in the $m_{\pi\pi}$ vs. $\cos\theta_{\pi\pi}$ plane, where $\cos\theta_{\pi\pi}$ is the angle between the kaon and the oppositely charged pion, calculated in the rest frame of the $\pi\pi$ system.
The different phases give rise to very different patterns of interference.

The ability to factorise the isobar coefficients and the resonance dynamics gives rise to great speed benefits if the parameters of the resonance model are fixed in the fits to data.
The integrals required to normalise the PDF need only be calculated once under these circumstances and the cached values can then be used in all iterations of the fits.
Similarly, the values of the resonance dynamics can be calculated once for each event in the data sample and these values cached.
In early versions of the Laura++ software, it was not possible to float resonance parameters since this caching mechanism was employed strictly, i.e. no recalculations were allowed.
However, recently a new caching and bookkeeping system has been developed in order to minimise the number of calculations that need be performed at any given iteration of the fit.
As such it is now possible (from v3r0 onwards) to float resonance parameters, although there is still some associated time penalty for such fits.

In extension to the isobar model there are various approaches implemented for modelling the S-wave, including the K-matrix~\cite{Chung:1995dx} and quasi-model-independent methods~\cite{Aitala:2005yh}.
An ongoing dialogue with theory colleagues will hopefully yield further improvements in this area in the near future.
One avenue currently being explored is the use of Veneziano models~\cite{Szczepaniak:2014qca}.

\section{MODELLING EXPERIMENTAL EFFECTS}

When analysing experimental data one must also account for presence of background processes and the acceptance of the experiment.
The variation of both the frequency of backgrounds and the signal reconstruction efficiency over the phase space can be provided to Laura++ in the form of two-dimensional histograms.
These can be either in the conventional Dalitz-plot coordinates or in the so-called ``square Dalitz plot''~\cite{Aubert:2005sk}.
They can be used as they are or interpolation methods (either linear or using cublic splines) can be applied to smooth them.
It is also possible to apply vetoes to particular regions of the phase space, e.g. to remove a peaking background, and to model the migration of mis-reconstructed signal events.
Furthermore it is possible to include variables other than the Dalitz-plot coordinates in the maximum likelihood fit in order to improve the discrimination between signal and the various background categories.
Many different PDF shapes are implemented in Laura++ to describe the non-DP variables, including forms that allow for correlations between the shape parameters and the DP position.
This feature was used extensively in the recent analysis of $B^+ \to K^0_\mathrm{S} \pi^+ \pi^0$ by the BaBar collaboration~\cite{Lees:2015uun}.

\section{ADVANCED FEATURES}

There are models that can be used to fit for $C\!P$ violation in decay, which have been used in the analyses of Refs.~\cite{Aubert:2008bj,Aubert:2009av,Lees:2015uun}.
A work in progress is to update the time-dependent models, used to cross-check the main fit in Ref.~\cite{Aubert:2009me}, to be more general.
In particular, it is of interest to allow for a non-zero difference of widths ($\Delta\Gamma$) of the mass eigenstates, to account for uncorrelated production environments such as $pp$ collisions at the LHC and to include the effects of variation of acceptance with decay time.
Once completed this will be made public in an upcoming release.

\begin{figure}[tb]
  \centerline{
  \includegraphics[width=0.45\textwidth]{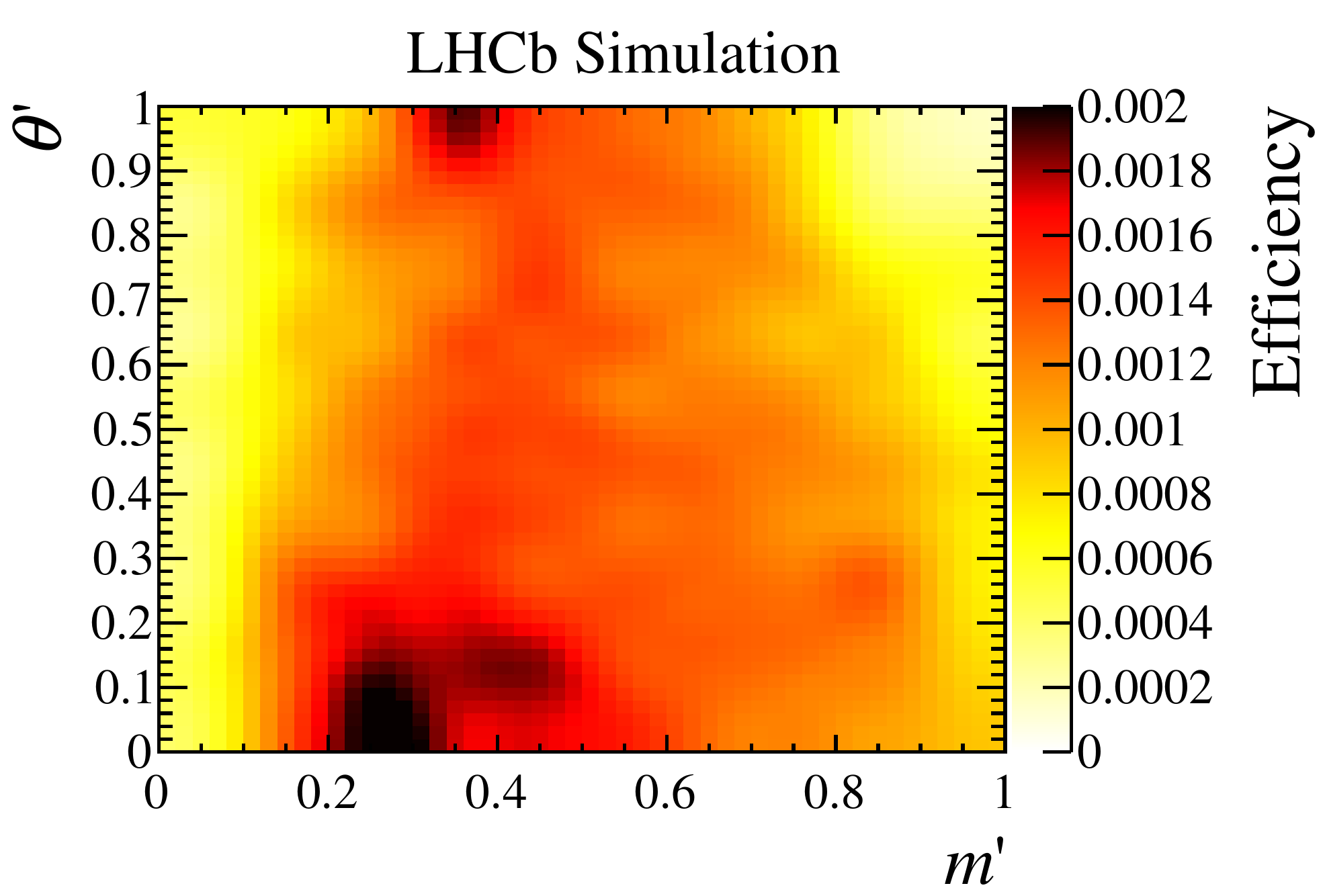}
  \includegraphics[width=0.45\textwidth]{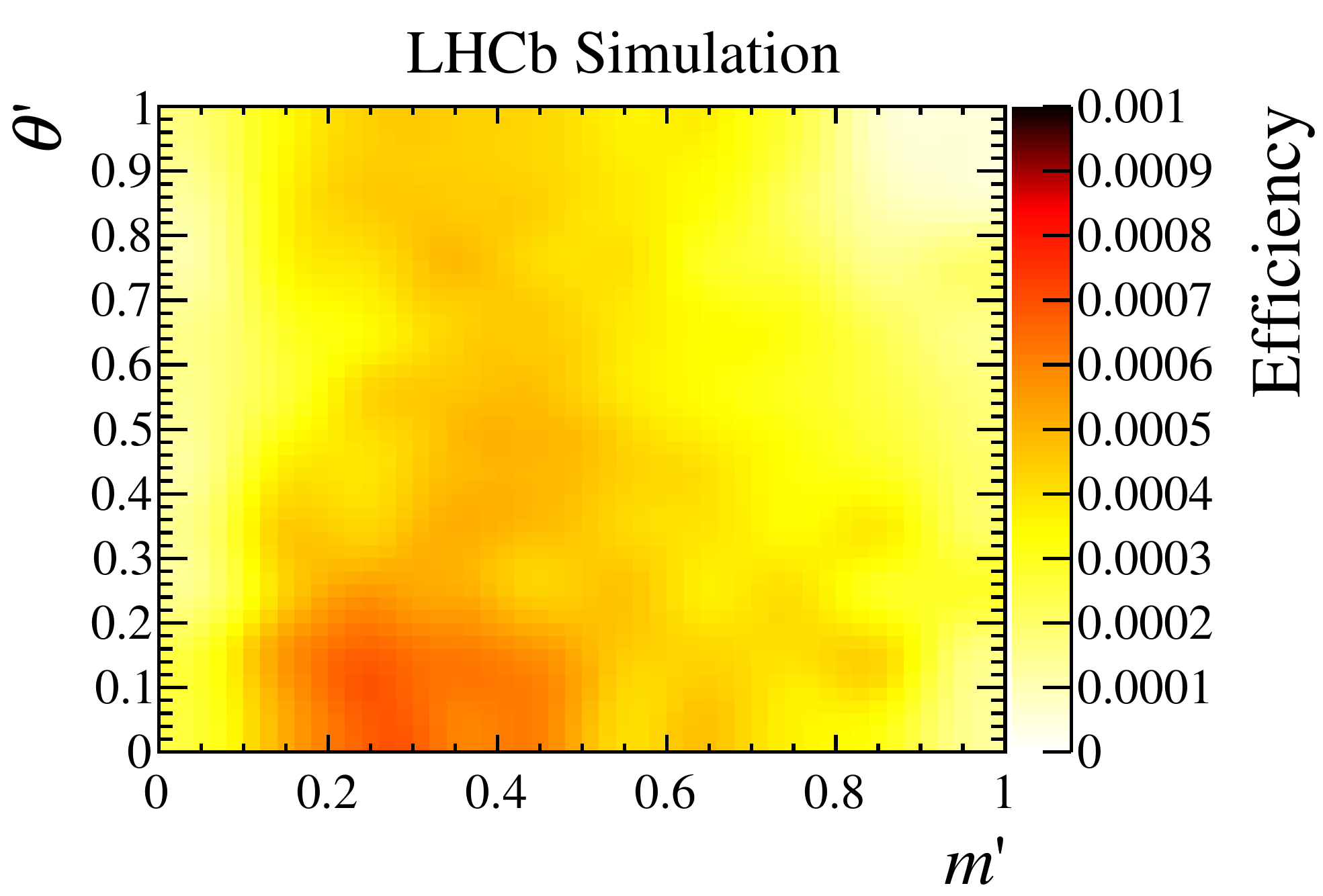}
  }
  \caption{
  Efficiency variation over the phase space for two different trigger categories in the recent LHCb analysis of $B^+ \to D^- K^+ \pi^+$. Figures from Ref.~\cite{Aaij:2015vea}.
  }
  \label{fig:tistos}
\end{figure}

A recently added feature is an implementation of the \textbf{\textit{J}}\hspace{-0.10em}\textsc{fit} framework~\cite{Ben-Haim:2014afa} to perform simultaneous fits to multiple data samples.
Originally devised for performing joint fits of data from different experiments, this framework is also extremely useful for many other purposes, such as handling different reconstruction or trigger categories that can give rise to different efficiency or background distributions over the Dalitz plot.
Figure~\ref{fig:tistos} shows the efficiency variation over the phase space for the two different trigger categories used in the recent LHCb analysis of $B^+ \to D^- K^+ \pi^+$~\cite{Aaij:2015vea}, which made use of this simultaneous fitting feature.
In addition it could be used for performing coupled-channel analyses or for fitting simultaneously to multiple decay modes in order to extract $C\!P$ violation observables following methods such as those proposed in Refs.~\cite{Ciuchini:2006kv,Gronau:2006qn,Latham:2008zs,Gershon:2008pe}.

\section{SUMMARY}

An overview of the Laura++ software package for performing Dalitz-plot analyses has been presented.
It is a mature package that has been used in many published analyses, mainly from the BaBar and LHCb collaborations.
It is under active development, with a number of new features added recently and others in the pipeline.
The package is open source and available for download from the HepForge website~\cite{HepForge}.
The website contains documentation to enable new users to get started quickly with the examples included with the package.
There are also mailing lists that allow questions to be asked to the developers and to obtain announcements of new releases.

\section{ACKNOWLEDGMENTS}
Laura++ has been developed with support from the University of Warwick, the Science and Technology Facilities Council (United Kingdom) and the European Research Council under FP7.
The authors of the package (Thomas Latham, John Back and Paul Harrison) would like to thank the following people for their invaluable contributions to the development and documentation of the package:
    Sian Morgan,
    Tim Gershon,
    Pablo del Amo Sanchez,
    Jelena Ilic,
    Eugenia Puccio,
    Mark Whitehead,
    Daniel Craik,
    Rafael Coutinho,
    Charlotte Wallace,
    Juan Otalora,
    Adlene Hicheur.
Many thanks also go to numerous members of the BaBar and LHCb collaborations for their helpful input.


\nocite{*}
\bibliographystyle{aipnum-cp}%
\bibliography{references}%

\end{document}